\documentclass[preprint2]{aastex}

\shorttitle{Long-time-series observations from Dome C}
\shortauthors{Rauer et al.}

\begin{document}

\title{Prospects of long-time-series observations \\from Dome C for transit search}
\author{Heike Rauer, Thomas Fruth and Anders Erikson}
\affil{DLR Institute for Planetary Research, Rutherfordstrasse 2, 12489 Berlin, Germany}

\begin{abstract}
The detection of transiting extrasolar planets requires high-photometric quality and long-duration photometric stellar time-series. In this paper, we investigate the advantages provided by the Antarctic observing platform \mbox{Dome C} for planet transit detections during its long winter period, which allows for relatively long, uninterrupted time-series. Our calculations include limiting effects due to the Sun and Moon, cloud coverage and the effect of reduced photometric quality for high extinction of target fields. We compare the potential for long time-series from \mbox{Dome C} with a single site in Chile, a three-site low-latitude network as well as combinations of \mbox{Dome C} with Chile and the network, respectively. \mbox{Dome C} is one of the prime astronomical sites on Earth for obtaining uninterrupted long-duration observations in terms of prospects for a high observational duty cycle. The duty cycle of a project can, however, be significantly improved by integrating \mbox{Dome C} into a network of sites.
\vspace{1cm}
\end{abstract}

\section{Introduction}
Extrasolar planets are currently detected via indirect methods. The radial velocity technique has proved most successful in recent years. However, photometric measurements of planets transiting in front of their central star through the line-of-sight to Earth have already provided more than 30 planet detections in the past years\footnote{See www.exoplanet.eu}. Many more transiting planet detections are expected in the very near future by the space missions CoRoT \citep{corot} and Kepler \citep{kepler}. Combining both methods allows the derivation of planetary radii, true masses and thus, the mean density of extrasolar planets. Transiting planets are therefore of particular importance to investigate basic planetary parameters.

The detection of extrasolar planets via the transit method requires long duration, high photometric quality lightcurves of a large number of stars. The major difficulties for ground-based transit searches are interruptions of the time series, e.g. due to the diurnal cycle or weather conditions, and the Earth's atmosphere limiting the photometric signal-to-noise-ratio \citep{transit_method}. Located on the Antarctic Plateau at $75^\circ$ southern latitude, \mbox{Dome C} provides the potential for long, uninterrupted high-quality photometric time series observations from ground during its winter period.

The maximum number of astronomical nights at \mbox{Dome C} in comparison to Mauna Kea has already been calculated by \citet{kenyon}, including the influence of the Sun, the Moon and cloud coverage. While \mbox{Dome C} has less astronomical nighttime, the percentage of cloud free observing hours is similar to Mauna Kea. Kenyon \& Storey do not discuss optimizations for long time series observations of target fields. Their paper, however, presents a detailed discussion of the expected sky brightness at \mbox{Dome C}. They estimate that atmospheric scattering should be exceptionally low, and even the sky brightness due to the Moon should be reduced. \mbox{Dome C} has reduced zodiacal light, no light pollution and only minor disturbances due to aurora. Furthermore, scintillation noise as a major limiting factor of photometric quality is expected to be about 3.6 times lower at Dome C compared to Chilean sites for long integrations \citep{kenyon2}, which is particularily important for the search of small planets. \mbox{Dome C} may therefore be the best site on Earth for stable high-accuracy photometry.

Alternatively, networks distributed favorably in longitude can also provide high duty cycles. A well known example is the Global Oscillation Network Group (GONG) for asteroseismological studies which include six network sites (four in the northern and two in the southern hemisphere). An estimate of the expected duty cycle for spectroscopic asteroseismology from \mbox{Dome C} in comparison to GONG has been presented by \cite{mosser}. They conclude that \mbox{Dome C} provides a better or similar performance than GONG for observing periods up to 100 days, attaining a duty cycle of up to 87\%. The network performs better for observing periods of half a year or more, but with a duty cycle of only about 77\%.

In this paper, we include the factors already investigated by Mosser \& Aristidi, like the Sun and cloud coverage. In addition we include the Moon and the varying extinction of the target fields during observations. Furthermore, we discuss not only a comparison of \mbox{Dome C} to low-latitude networks, but also a combination of \mbox{Dome C} with other sites. A comparison between \mbox{Dome C} and the CoRoT mission has already been presented \citep{ARENA2}. The resulting potentials for long-duration time-series photometric observations are then discussed in the context of maximizing the probability to detect extrasolar photometric transits.

\section{Duty Cycle Simulation Procedure}\label{model}
Observations from the ground are limited for various reasons. The main restriction is imposed by the Sun for observations in the visible spectral range. Dark time can be reduced further by periods of full moon, because during such periods the sky background noise level reduces the photometric accuracy. Selected target fields move across the sky, rising and setting and thus further limiting any observation campaign from a single site. After a brief discussion of the underlying time-frame in \S\ref{period}, the actual determination of observation times from these astronomical constraints will be the subject of section \S\ref{geometricalconstraints}. In practice, local weather might likewise limit observations, e.g. due to cloud coverage. Section \S\ref{clouds} describes the model used for simulating this effect. Finally, the total observing time on a chosen target field as well as the photometric quality obtained crucially depend on its coordinates. In \S\ref{targetoptimization} we therefore present a simple model designed to provide a quantitative solution to the target field selection process.

\subsection{Period of Observations}\label{period}
The selected time-frame for our simulation limits possible observations to a certain period and therefore affects the results systematically. The most appropriate choice is to consider a one year period for each site studied, as the Earth's orbit around the Sun has the most important impact on observability. After one year has elapsed, time intervals of possible observations would basically show again the same pattern for any given target field. The results presented here correspond to one year ($365.24d$), starting on 1 March 2008, 0:00h UT.

Furthermore, only one field is considered per site to determine the maximum possible time coverage for planet detections. Thus, planet detections at large orbital distance was given highest priority, rather than optimizing for detections of very short-period planets by maximizing the number of fields.

\subsection{Astronomical Constraints}\label{geometricalconstraints}
The Sun and Moon affect the background noise level depending on their position relative to the observer. In order to decide whether a target is observable at a certain time or not, some criteria are needed. The most common and efficient way is to limit the three local altitudes $h$, i.e. to apply
\begin{eqnarray}\label{limit_h}
 h_\odot       &\leq & h_\odot^{max}  \hspace{.5cm} \mathrm{for\ the\ sun,} \nonumber\\
 h_{Moon}      &\leq & h_{Moon}^{max} \hspace{.32cm} \mathrm{for\ the\ Moon\ and}\\
 h_*           &\geq & h_*^{min}      \hspace{.55cm} \mathrm{for\ the\ target.}\nonumber
\end{eqnarray}

For example $h_\odot^{max}=h_{Moon}^{max}=0$ calculates observation times without direct sun- or moonlight, as they would be restricted to stay always below the horizon. However, since scattered light might likewise disturb observations significantly, the common limit $h_\odot^{max} = h_{Moon}^{max} = -8^\circ$ was used in this work, at which the contamination of the photometry is negligible. Furthermore, target fields below $30^\circ$ altitude are not considered observable due to their poor photometric quality, i.e. $h_*^{min} = 30^\circ$.

Moreover, the Moon's contribution to sky brightness is a function of its phase $\phi_{Moon}$. Therefore the Moon criteria in eq. (\ref{limit_h}) was relaxed by additionally allowing for situations corresponding to
\begin{equation}\label{limit_moonphase}
 \left( h_{Moon} > h_{Moon}^{max} \right) \wedge \left( \phi_{Moon} \leq \phi_{Moon}^{max} \right),
\end{equation}
i.e. times when the Moon altitude is above its limit, but the corresponding phase stays below a specified parameter $\phi_{Moon}^{max}$. Furthermore, the distance between Moon and target field can be restricted to stay above a certain value, that is
\begin{equation}\label{limit_moontarget}
 d_{Moon} \geq d_{Moon}^{min},
\end{equation}
where $d_{Moon}$ denotes the angle between Moon and target on their great circle. Based on our observational experience, these last two limits were chosen to be $\phi_{Moon}^{max} = 0.9$ and $d_{Moon}^{min} = 20^\circ$ at which the influence on photometric accuracy starts to be significant.

Compared to low-latitude sites, the sky background is expected to be darker at Dome C for the same local positions of Sun, Moon and target due to the low aerosol content of the air \citep{kenyon}. Nevertheless, we refrained from adjusting the limits as this effect is still to be examined more quantitatively.

\subsection{Clouds}\label{clouds}
The average cloud pattern at a given site was simulated using a model by \citet{Hill1985}, which requires two input parameters, $\rho$ and $\tau$, derived from weather statistics. $\rho$ stands for the average fraction of clear skies, whereas $\tau$ denotes the mean duration of cloudless periods. Both can vary seasonally.

According to the model, a simple pattern is created that consists of periods that either allow observations (clear sky) or not (clouds) for any given site and season. The duration of a \textit{single} cloudless period is governed by a gamma distribution of the form
\begin{equation}\label{clouds_p(t)}
 p(t)=\frac{4t}{\tau^2}\,e^{-2t/\tau},
\end{equation}
so that $p(t)$ denotes the probability to find an uninterrupted interval of length $t$. In order to meet the given fraction $\rho$ of clear sky, the mean duration of cloudy periods is then clearly given by $\tau (1-\rho)/\rho$. Substituting this value for $\tau$ in eq. (\ref{clouds_p(t)}), one obtains the corresponding probability distribution of cloudy weather. An alternating sequence of clear and cloudy periods will therefore provide an approximation of real sky coverage.

So, periods of potential observations (calculated using the geometrical constraints) are further reduced by only selecting clear sky times. As these are based upon random processes, 500 Monte Carlo runs are performed for each configuration in order to achieve statistically significant results.

\subsection{Target Optimization}\label{targetoptimization}
For the purpose of site comparisons, the calculation of observational time series is often derived from astronomical night times only, i.e. without including a specific target field. These results form an upper limit on the duty cycle, but are usually not achieved by any \textit{single} target. For instance, most fields can only be observed for a fraction of a year due to the Earth's annual motion around the Sun. However, the outcome of transit search programmes largely relies on how much time a chosen field is being observed. In particular, only long time-series might provide a sufficient detection rate for planets with Earth-like orbits. Including target field coordinates into the simulation is therefore necessary for an accurate analysis of duty cycles.

To select optimized target field coordinates, we posed the question: For a given site and time interval, which target field will provide the longest time coverage? Two parameters are relevant here:
\begin{itemize}
 \item Observation time $T$
 \item Mean Airmass $\overline{X}$
\end{itemize}
Considering the observation time, fields observable over long periods are favoured. However, the photometric quality is reduced with increasing airmasses. Therefore our calculations seek to minimize the mean airmass.

For that, both parameters are calculated as a function of potential equatorial target coordinates $(\alpha,\delta)$ throughout the whole sky for each site. In order to optimize them simultanously, they must subsequently be weighted against each other by defining a function $\Theta \left(T(\alpha,\delta),\overline{X}(\alpha,\delta)\right)$. Using such a weighting mechanism then provides a sky map from which we can easily derive a maximum value.

\begin{figure}[htc]
 \plotone{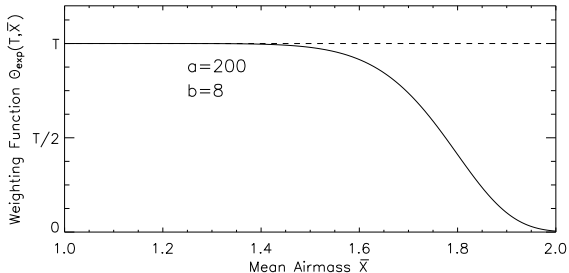}
 \caption{airmass weighting function, see eq. (\ref{Theta})}
 \label{fig:am_weighting}
\end{figure}

Although the choice of a specific function $\Theta$ is somewhat arbitrary, it should in general show a reasonable dependency on its two variables $T$ and $\overline{X}$: Firstly, $\Theta$ should be monotonically decreasing in $\overline{X}$, as the photometric quality can only be reduced with larger airmasses. Secondly, only a linear dependency on the observing time $T$ is meaningful. Finally, we have already excluded target fields with airmasses $X>2$ by setting $h_*^{min}=30^\circ$, thus it is consequent to demand $\Theta\rightarrow 0$ for $\overline{X}\geq 2$.

These requirements are well fulfilled by the following approach:
\begin{equation}\label{Theta}
 \Theta_{exp}\left(T,\overline{X}\right)=\frac{T}{a^{(\overline{X}-1)^b}}
\end{equation}
The parameters $a=200$ and $b=8$ have been chosen to reproduce the dependency of photometric accuracy on the airmass of observations, as illustrated in Fig. \ref{fig:am_weighting}: In the region $1.0\leq \overline{X} \leq 1.4$ we get $\Theta \approx T$, which accounts for the usually negligible variation of the signal-to-noise ratio of photometric data recorded at these airmasses. Beyond that the photometric quality is significantly reduced, resulting in a strong devaluation of observing times with an average airmass close to 2.

This procedure yields an efficient choice of target coordinates for site networks. We then maximized the total time of observation from at least one site and decided to observe from whichever site provided the smallest airmass whenever simultaneous observations were possible.

Target coordinates found by this procedure obviously take into account astronomical configurations only. Additional factors are required to optimize the field for an exoplanet transit search programme. In particular, the target field must contain a large number of dwarf stars. The required high number has to be weighted against crowding effects in dense star fields. Therefore, this optimization step depends on the FOV and the spatial scale of the instrument chosen. It has to be performed for a particular instrument and is therefore not within the scope of this paper which concentrates on the astronomical boundary conditions in terms of optimized duty cycle.

\section{Input Parameters}\label{parameters}

\subsection{Geographic Coordinates}
As already mentioned, we aim to compare the window function obtained from Antarctica with one or more low-latitude sites as well as a combination of \mbox{Dome C} with such locations. In addition to \mbox{Dome C}, three sites with similar low-latitudes are considered for a potential southern hemisphere network. The location of the BEST II telescope (Erikson \textit{et al.}, in prep.) at Cerro Armazones was chosen in Chile. In Africa a site next to the Gamsberg in Namibia was included in our calculations, whereas the Anglo-Australian Observatory (AAO) was  selected on the Australian continent. The geographical coordinates are listed in Table \ref{localcoordinates}.
\begin{table}[htc]\footnotesize
\caption{Local coordinates of considered sites}
\label{localcoordinates}
\begin{tabular}{llll}
  \hline\hline
  Site    & Longitude & Latitude & Altitude \\\hline
  Dome C  & E $123^\circ 23'$      & S $75^\circ 06'$      & 3260m \\
  BEST II & W $70^\circ  11' 35''$ & S $24^\circ 35' 24''$ & 2840m \\
  Namibia & E $16^\circ  06' 38''$ & S $23^\circ 17' 25''$ & 1450m \\
  AAO     & E $149^\circ 03' 52''$ & S $31^\circ 16' 24''$ & 1165m \\\hline
\end{tabular}
\end{table}

\subsection{Weather Data}
The two parameters used by the weather model, $\rho$ and $\tau$, can exhibit strong seasonal and geographical variability. For the simulation of free-sky windows, monthly mean values are used for each site in Table \ref{localcoordinates}.
\begin{figure}[htc]
 \plotone{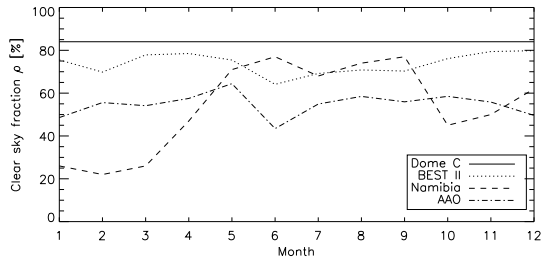}
 \caption{Monthly average values of the clear sky fraction $\rho$ for Antarctica, Chile, Namibia and Australia}
 \label{fig:weather_compare} 
\end{figure}
Figure \ref{fig:weather_compare} shows the dependency of the assumed average clear sky fraction $\rho$ on time and location. For \mbox{Dome C}, winter observations from \citet{mosser} with mean values of 6.8 consecutive clear days and 84\% overall clear skies were extrapolated to the whole year. In the case of Chile, cloudiness data from the European Southern Observatory (ESO) Paranal\footnote{See www.eso.org/gen-fac/pubs/astclim/paranal/clouds/} were averaged for 1983-2006, whereas publicly available \textit{surface synoptic observations} (SYNOP)\footnote{See www.ogimet.com} from 2004-2007 were used for the AAO. Finally, an ESO site survey provided values for $\rho$ at close-by Gamsberg in Namibia \citep{gamsberg}.

For Australia, the average duration of clear periods in the data set was found to be $\tau=0.5d$. For Chile and Namibia we assumed $\tau = 5d$ according to \citet{Hill1985}. However, the parameter $\tau$ only affects the duration of individual observations and is thus of minor importance. For example, using values $1 \leq \tau \leq 10$ for Chile does neither change the total observation time $T$ nor the time coverage of planetary transits.

\section{Simulations and Results}\label{results}

We first validated the model against previous studies.
Calculated celestial coordinates agree with results from the publicly available \textit{JPL Horizons}\footnote{http://ssd.jpl.nasa.gov/?horizons} ephemeris software to within 30 arcseconds for the years 1900-2100. In addition to this, the results of \citet{kenyon} for available dark time at \mbox{Dome C} and Mauna Kea were confirmed when we adopted their slightly differing limits for sun- and moonlight.

Then, optimal target fields have been simulated for all single sites as well as for three possible combinations of the sites. The various configurations are summarized in Table \ref{tab:results}.

\subsection{The Effect of Duty Cycle Limitations}
\begin{figure}[htc]\centering
 \hspace{-7mm}
 \plotone{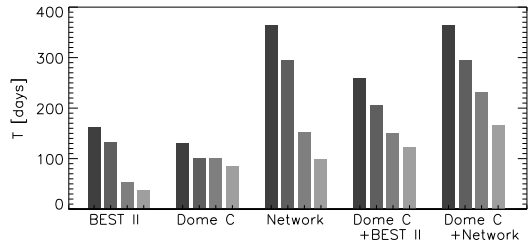}
 \vspace{-2mm}
 \caption{Total observing time $T$ for one year from different sites, broken down into limiting effects. The following restrictions were gradually included (from dark to light bars): (1) sunlight (2) moonlight (3) one target field only (see Table \ref{tab:results}) (4) clouds}
 \label{fig:effects}
\end{figure}
As a first step, we investigated the limitations imposed on time-series observations by each of the constraining factors considered individually. Figure \ref{fig:effects} shows calculated values of $T$ taking into account firstly only sunlight, then gradually including moonlight, target field visibility and finally cloud coverage.

At a single site like BEST II, the maximum dark time is 163 days per year due to the diurnal cycle ($h_\odot^{max} = -8^\circ$, dark bars in Fig. \ref{fig:effects}). At \mbox{Dome C}, the dark time per year is somewhat less (135 days) due to its high latitude. On the contrary, a low-latitude network can in principle observe all year round. However, effects like moonlight, target field selection and clouds further reduce the maximum available observing time.

The target field selection does not impose a strong restriction on available observing time from \mbox{Dome C}, because many fields are above $h_*^{min}$ almost continuously during the polar night. At low latitude sites, the full available dark time can only be filled with observations of several target fields over the year. For one field it is, of course, significantly reduced (compare third bars from left for each site in Fig. \ref{fig:effects}). 

In addition, there are also severe effects by clouds and the final performance is therefore significantly reduced in comparison to the ideal case (fourth bar). Due to the unequal impact of limitations upon different sites it is therefore necessary to consider all four effects simultaneously.

\subsection{Duty Cycles for one field per year}

\begin{deluxetable}{lcccrrrcc}
 \tabletypesize{\footnotesize}
 \tablewidth{0pt}
 \tablecaption{Optimized target coordinates (J2000.0)\label{tab:results}}
 \tablehead{
  \multicolumn{1}{l}{Site(s)} & \colhead{$\alpha$} & \colhead{$\delta$} & \colhead{Clouds} & \colhead{T} & \colhead{$t_{mean}$} & \colhead{$t_{max}$} && \colhead{$\overline{X}$}
 }
 \startdata
  Dome C           & $00^h 00^m$ & $-90^\circ 00'$ & without & $101.2d$ & $13.3h$ & $ 23.0h$ && 1.03 \\
                   &             &                 & with    & $ 85.0d$ & $11.2h$ & $ 15.3h$ && 1.03 \\\cline{4-9}
  BEST II          & $18^h 47^m$ & $-47^\circ 06'$ & without & $ 53.0d$ & $ 5.1h$ & $  9.6h$ && 1.33 \\
                   &             &                 & with    & $ 38.0d$ & $ 4.6h$ & $  6.1h$ && 1.41 \\\cline{4-9}
  Network          & $18^h 41^m$ & $-51^\circ 42'$ & without & $151.2d$ & $ 6.6h$ & $551.5h$ && 1.33 \\
                   &             &                 & with    & $ 98.4d$ & $ 5.4h$ & $ 15.1h$ && 1.40 \\\cline{4-9}
  Dome C + BEST II & $06^h 45^m$ & $-57^\circ 06'$ & without & $150.0d$ & $ 9.4h$ & $ 23.0h$ && 1.33 \\
                   &             &                 & with    & $122.5d$ & $ 8.9h$ & $ 13.3h$ && 1.34 \\\cline{4-9}
  Dome C + Network & $05^h 29^m$ & $-56^\circ 17'$ & without & $231.4d$ & $ 8.6h$ & $ 23.0h$ && 1.33 \\
                   &             &                 & with    & $165.4d$ & $ 7.6h$ & $ 15.6h$ && 1.34 \\
 \enddata
 \tablecomments{Results of field optimization are shown for single sites at \mbox{Dome C} and Chile (BEST II) as well as potential combinations of southern hemisphere telescopes (Network = BEST II + Namibia + AAO). The sum of all observing times on the corresponding field is given by $T$, whereas $t_{mean}$ and $t_{max}$ denote the mean and maximum lengths of \textit{single} periods of continuous observations, respectively. Photometric quality was taken into account by calculating the mean airmass $\overline{X}$ during potential observations.}
\end{deluxetable}

\begin{figure}
 \hspace{7mm} \epsscale{0.8}
 \plotone{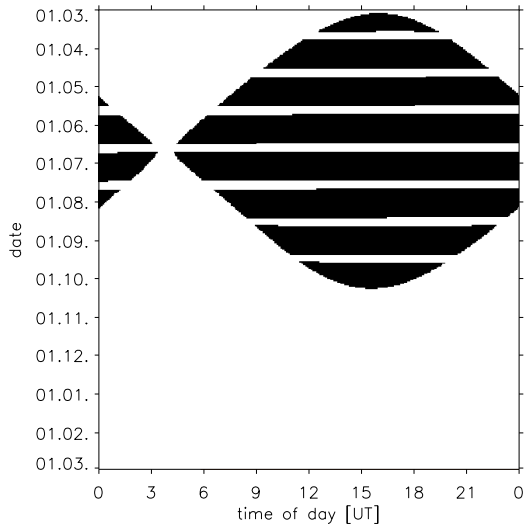}
 \vspace{5mm}
 \caption{Window function for observations from \mbox{Dome C} (2008): Black areas mark times of possible observations.
   The effects of sun- and moonlight are clearly visible; the chosen target (Zenith) is always observable, clouds are not included.}
 \label{fig:domec}
\end{figure}

\begin{figure}
 \hspace{7mm} \epsscale{0.8}
 \plotone{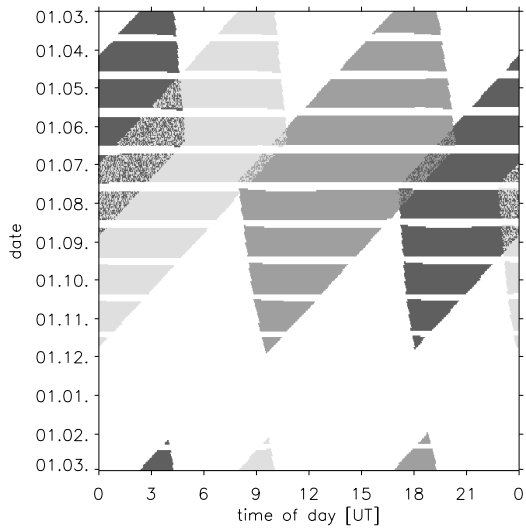}
 \vspace{5mm}
 \caption{Window function for a southern network of BEST II \textit{(light gray)}, Namibia \textit{(medium gray)} and Australia \textit{(dark gray)} without clouds. The field $(18^h 41^m; -51^\circ 42')$ is continously observable during the weeks around winter solstice.}
 \label{fig:network1}
\end{figure}

In Table \ref{tab:results} potential target coordinates $(\alpha,\delta)$ with maximum $\Theta_{exp}\left(T,\overline{X}\right)$ are given for each combination of sites. Also the corresponding values of total observation time $T$ and mean airmass $\overline{X}$ are given -- with and without clouds, respectively. Furthermore, the mean and maximum duration of single continuous periods of observations is shown as $t_{mean}$ and $t_{max}$.

The window function itself is illustrated in Fig. \ref{fig:domec} for \mbox{Dome C}. The long observing periods during the polar night are clearly visible. However, continuous observations do not exceed 23 hours due to the sun rising above $-8^\circ$ at noon even around winter solstice. Nevertheless, the task of observing the \textit{same} target from a single site as long as possible can best be fulfilled from Antarctica. Even when weather statistics are taken into account, \mbox{Dome C} clearly outcompetes even an excellent low-latitude site such as BEST II. Under realistic conditions, the mean total duration of observations of a target field with cloud coverage is 85 days from Antarctica which is more than twice as long as from Chile (38 days).

In order to achieve a similar performance without \mbox{Dome C}, the combination of at least three low-latitude sites would be necessary. Figure \ref{fig:network1} shows the performance of a network consisting of Chile, Namibia and Australia. This would in principle allow for very long continuous observations taken in turns, peaking in mid-winter at a maximum period close to one month between full moons. But interruptions due to cloud coverage limit long runs and reduce the total observation time to 98 days. However, the network would still yield about two weeks more than \mbox{Dome C} in terms of observing duration.

\begin{figure*}
 \hspace{2mm}
 \begin{minipage}[b]{0.48\linewidth}
   \hspace{7mm}
   \plotone{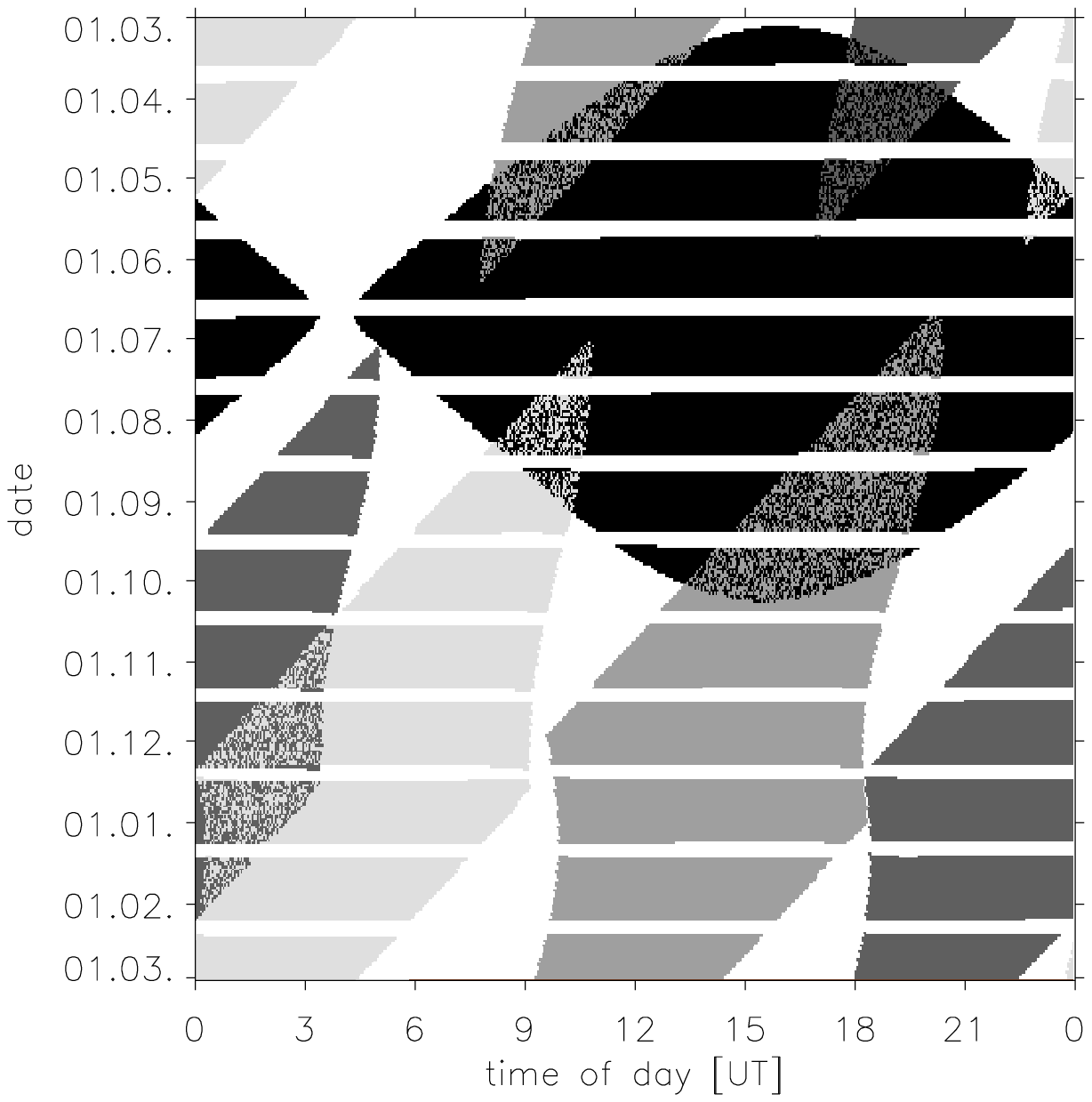}
 \end{minipage}
 \begin{minipage}[b]{0.48\linewidth}
   \hspace{7mm}
   \plotone{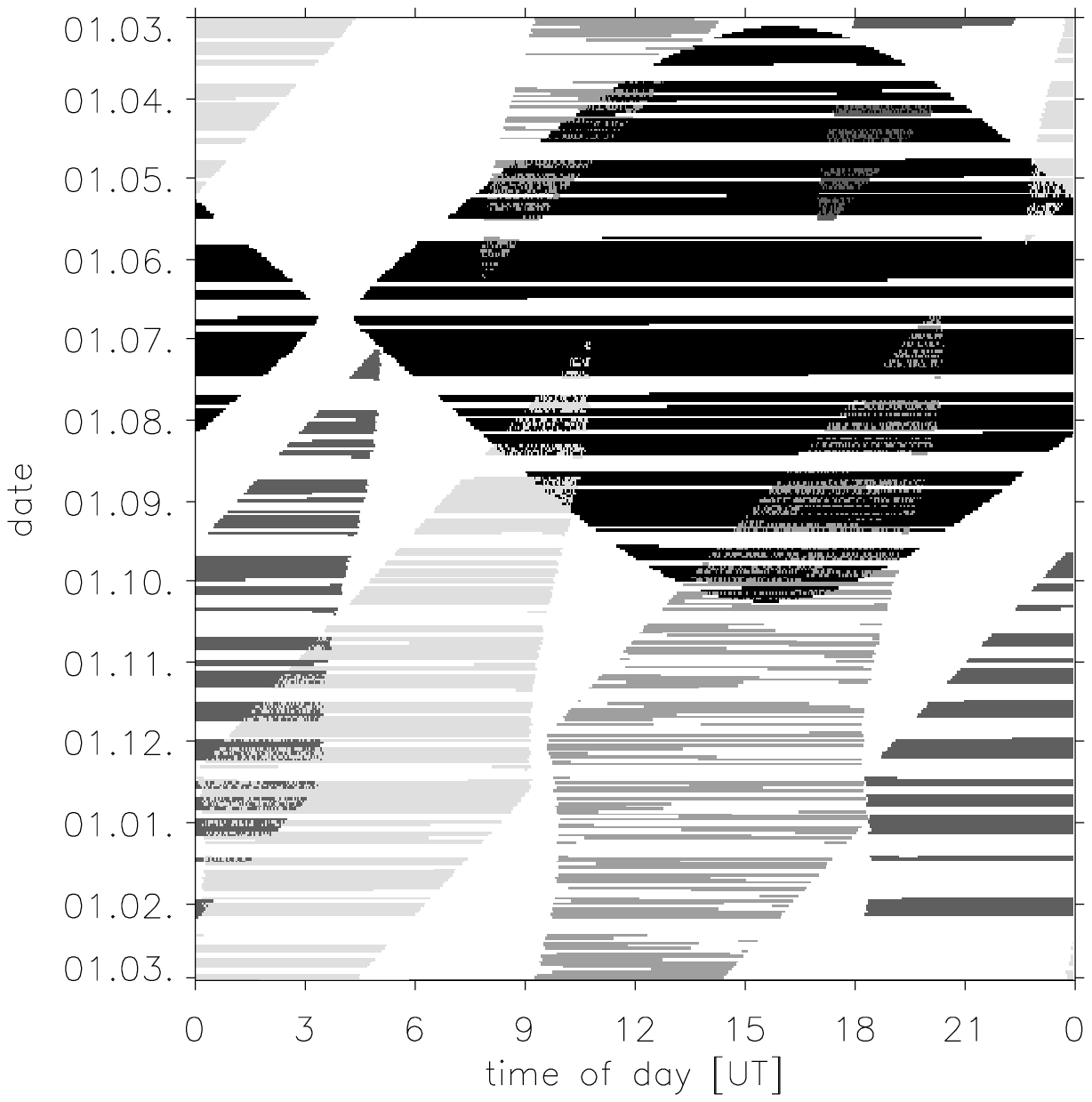}
 \end{minipage} \vspace{5mm}
 \caption{Possible observing times for a three site low-latitude network together with \mbox{Dome C} \textit{(colors as in Figs. \ref{fig:domec} \& \ref{fig:network1})}. The corresponding optimal target field $(05^h 29^m; -56^\circ 17')$ is located on the summer night sky. The left side shows the theoretical window function using geometrical constraints (Sun, Moon, target visibility) only, whereas weather statistics are taken into account on the right.}
 \label{fig:network2}
\end{figure*}
 
\begin{figure*}
 \centering
 \includegraphics[width=0.8\linewidth]{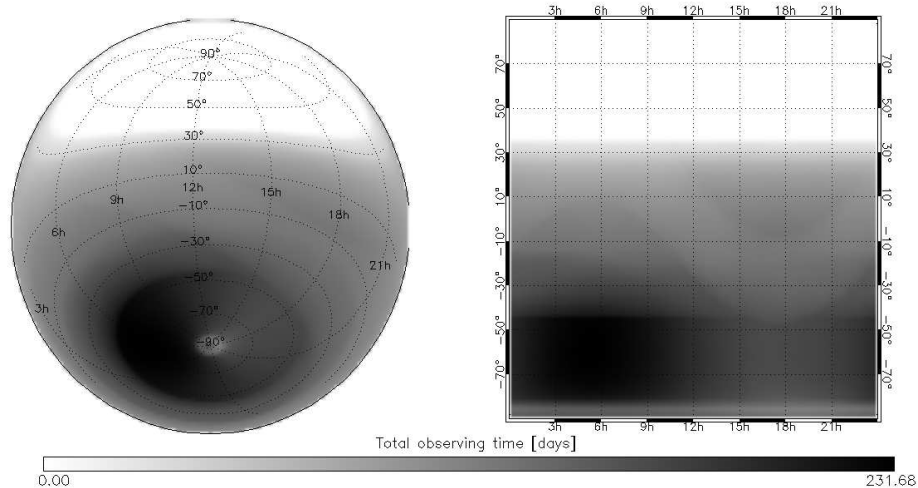}
 \caption{Total observing time as a function of target coordinates $(\alpha,\delta)$ obtained with a southern hemisphere network including Antarctica. The left half shows $T(\alpha,\delta)$ throughout the whole sky in a Lambert projection, whereas the same data is visualized on the right with equidistant spacing for each angular dimension. Cloud coverage is not included here.}
 \label{fig:target_plot}
\end{figure*}

Beyond the direct comparison of \mbox{Dome C} with one or more low-latitude sites, network configurations including Antarctica deliver very interesting prospects. As \mbox{Dome C} is restricted to observations during the winter months only, low-latitude sites can also monitor targets located on the summer night sky around $\alpha\approx 6^h$. With the additional choice of $\delta\approx -55^\circ$, such a field would always remain more than $30^\circ$ above the horizon at \mbox{Dome C}. The consequence is a significantly increased time statistics, as the overlap of observations from Antarctica with the other sites would be marginal. Figure \ref{fig:network2} illustrates the excellent duty cycle expected from a network consisting of all four considered sites including \mbox{Dome C}, whereas Fig. \ref{fig:target_plot} shows the functional dependency of the total observation time on field coordinates.
Combining only \mbox{Dome C} and BEST II, it would be possible to observe for a total of 123 days per year under realistic conditions, thus exceeding the duty cycle of the three-site low-latitude network by 25\%. Finally, increasing the total obervation time further to 165 days by including Namibia and Australia into the network would provide an outstanding performance for an Earth-bound project.

\subsection{Time Coverage of Planetary Transits}

\begin{figure*}\centering
 \hspace{0.08\linewidth}
 \includegraphics[angle=90,width=0.9\linewidth]{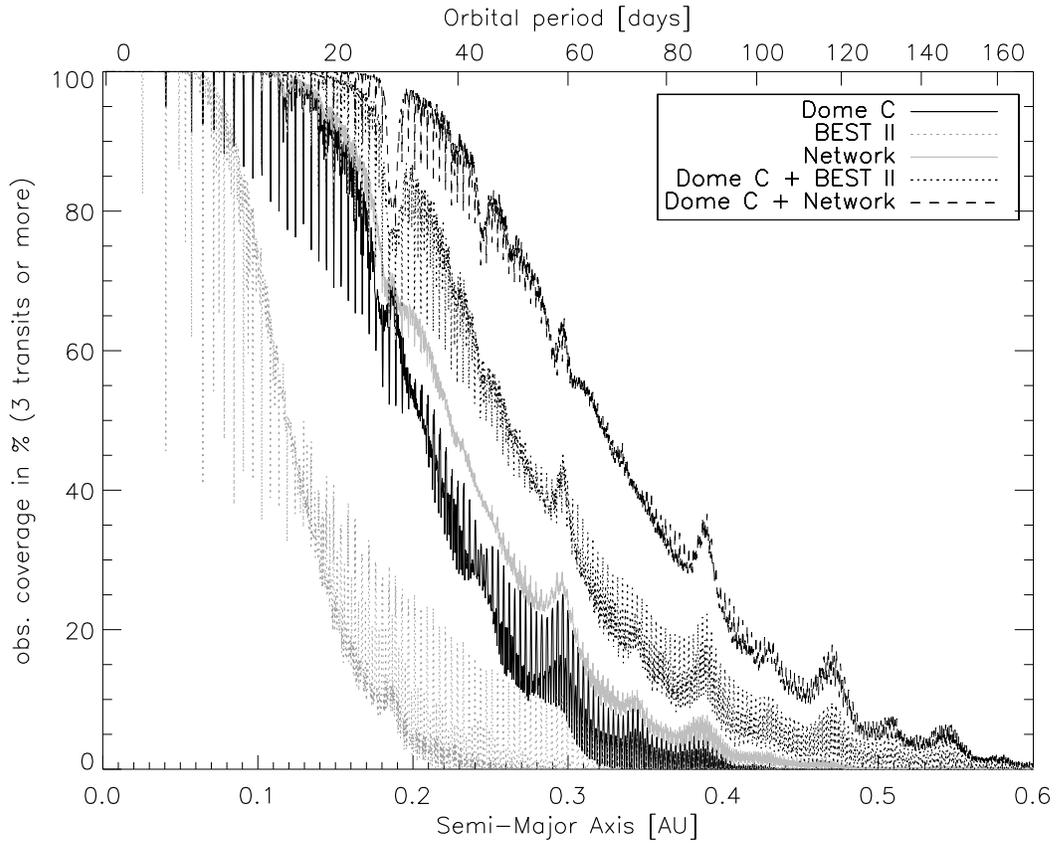}
 \caption{Observational coverage of at least three transit events, based on duty cycle calculations including cloud coverage for \mbox{Dome C} \textit{(dark solid line}), BEST II \textit{(light dotted line}) and a southern network of BEST II, Namibia and the AAO \textit{(light solid line}). In addition, the performance of \mbox{Dome C} combined with BEST II \textit{(dark dotted line}) is shown as well as combined with the whole southern network \textit{(dark dashed line}).}
 \label{fig:cov3_cld}
\end{figure*}

In the following, the efficiency of transit detection is examined. Figure \ref{fig:cov3_cld} shows a comparison of observational coverage based on the simulated window functions. Simulations have been performed as a function of the orbital period and for three or more transit events, which is usually the minimum requirement for detection. That is, a coverage of 100\% means at least three events of a potential transiting planet will be observed for a given orbital distance. Note that we only take into account the time sampling here, so the actual detection probabilities will be reduced further e.g. by the signal-to-noise ratio or geometric orientation.

All curves in Fig. \ref{fig:cov3_cld} show peaks and dips caused by interruptions in the data sampling e.g. due to the diurnal cycle, full moon or cloud coverage. Periodic data gaps lead to a reduced observational coverage for orbital periods which are multiple of interruption frequencies.

\mbox{Dome C} can provide continuous observational coverage for planets with up to about two weeks orbital period, extending to 30 days with a probability of 50\%. In contrast to that, it is very unlikely to detect planets with periods significantly longer than 15 days for a single low-latitude site like BEST II in Chile within one year. However, an almost equal performance to \mbox{Dome C} could be achieved by the combination of three low-latitude sites.

Combining a telescope at \mbox{Dome C} with one in Chile extends the full observational coverage to about three weeks and leads to a significantly improved performance for larger periods compared to \mbox{Dome C} alone.

Finally, the combination of \mbox{Dome C} with a network including Chile, Namibia and Australia could, in addition to a good coverage of small orbits, in principle still detect planets with periods of more than 60 days at a reasonable coverage.

\section{Conclusions}

We have investigated the prospects for long-duration uninterrupted photometric time series from \mbox{Dome C} and other low-latitude sites. Contrary to previous studies we \textit{simultaneously} took into account the limitations of the Sun and Moon, clouds and target field extinction variations. 

Due to the long winter, \mbox{Dome C} has an outstanding performance in observing a target field over a long time period from a single site. Furthermore, the effective observing time period can be extended if \mbox{Dome C} is combined with other sites which can provide data during the summer period. For example, the combination of \mbox{Dome C} with a site in Chile extends the observational coverage significantly. Combining \mbox{Dome C} with a low-latitude network of sites at Chile, Namibia and Australia provides prospects for observational duty cycles that can be improved upon only by observations from space. 

To predict quantitative transit detection rates, the photometric quality of \mbox{Dome C} must be investigated directly. However, for this we have to await the results of photometric site testing projects like the Antarctica Search for Transiting Extrasolar Planets (ASTEP, \cite{astep}) for a detailed study. Nevertheless, the extensively long observational time baseline for photometric variability studies clearly demonstrates the extremely encouraging prospects of this site in Antarctica.

\end{document}